**Fueling Peter's Mill: Mikhail Lomonosov's Educational Training in Russia and Germany, 1731–1741**

*Abbreviated title:* Mikhail Lomonosov

Robert P. Crease and Vladimir Shiltsev

**Fueling Peter's Mill: Mikhail Lomonosov's Educational Training in Russia and Germany, 1731–1741**

Robert P. Crease and Vladimir Shiltsev[*]

*Abstract:* This article, the second in a series about the Russian scientist Mikhail Lomonosov (1711–1765), traces his education from his arrival in Moscow in 1731 to study at the Slavo-Greco-Latin Academy, through his admission to the St. Petersburg Academy of Sciences in 1736, to his trip abroad to complete his educational studies from 1736 to 1741. Lomonosov's story during this time opens a vista on the introduction of modern physics and modern science into Russia. Michael D. Gordin has argued that Peter the Great's plans to Westernize Russia were more broadly conceived than he is usually credited, with ambitions that exceeded mere utilitarian and pragmatic goals. Lomonosov's career trajectory is a good example, illustrating how different aspects of the Petrine vision intersected with and reinforced each other. The article ends with Lomonosov's return to Russia from Germany in 1741, an important landmark in the growth of the Academy and of Russian science.

[*] Robert P. Crease (corresponding author) is Professor of Philosophy at Stony Brook University. Vladimir Shiltsev is Director of the Accelerator Physics Center at Fermilab.

**Introduction**

En route to Sweden in 1724, the Russian statesman and historian Vasily Tatishchev stopped at his patron Peter the Great's summer palace, the Peterhof, just outside St. Petersburg. There he was accosted by Lawrence Blumentrost (figure 1), an advisor to Peter who was about to become president of one of Peter's projects, the St. Petersburg Academy of Sciences, which was in the planning stages. Blumentrost asked Tatishchev, who was headed to Sweden to supervise the training of Russian youth in mining, to look for educated scholars there who might be recruited to teach at the new Academy. Tatishchev laughed and said that this was like asking him to build a magnificent water pump with no place to put it and no water. Hearing the remark, Peter asked Tatishchev to explain. The new Academy would be useless, Tatishchev replied, for in Russia there were no teachers nor anybody to be taught. Peter replied, "I have to grind vast quantities of wheat, but there are no mills and no water in sight. There is enough water in the distance, but no time left in my life to dig the channel to bring it to me. So I begin to build the mill anyway, with the channel only existing in my imagination, leaving it to my heirs to finish the construction that will get the water to the mill.… Though I will not see the fruits, I hope that these will come to my homeland."[1]

The story of Mikhail Lomonosov's educational experiences sheds light on how the flow of water to supply Peter's visionary mill began. Lomonosov (1711–1765) lived during a period of rapid artistic, cultural, linguistic, and scientific change in Russia, the pace and impact of which exceeded that of any Western country in this time (figure 2). Peter had initiated these changes, which affected nearly every part of Russian society. Not all of Peter's goals were realized, but his attempts to introduce European-style science and scientific institutions were part and parcel of a wider transformation in literature, poetry, education, and language. Lomonosov benefitted from and played a role in changes in all these areas. Michael D. Gordin has argued that Peter had more broadly conceived plans to Westernize Russia than he is usually credited with, involving more than utilitarian and pragmatic goals. Lomonosov's career trajectory is a good example, illustrating how different aspects of the Petrine vision intersected with and reinforced each other.[2] That trajectory took him from the Slavic-Greek-Latin Academy in Moscow, where he seemed destined to join the lower ranks of the clergy, to the St. Petersburg Academy of Sciences, where he was a student, adjunct, and eventually its first native Russian academician.[3]

**Moscow (1731–1735)**

Lomonosov was born on Kurostrov Island, in Kholmogory county about seventy-five kilometers from Arkhangelsk.[4] His father was a fisherman and during his early years Lomonosov learned the wide variety of practical skills that any fisherman in that frontier region had to acquire, from boatbuilding and repair to sailmaking, ropemaking, weather prediction, and knowledge of the kinds and habits of sea creatures. Thanks to his father, too, Lomonosov had numerous personal contacts both in and around the region. Despite its large size, the region's sparseness meant that survival depended on interdependence, and many inhabitants had either encountered each other or shared acquaintances.[5]

Highly ambitious and with a voracious intellectual appetite, Lomonosov was frustrated by the poor education he was receiving. In December 1730, the nineteen-year-old youth borrowed three rubles and a jacket and joined a trade convoy bound for Moscow. Lomonosov made the 1,170 km journey, partly on foot and partly on sleigh, in somewhere between twenty-two and thirty-seven days. When he arrived in Moscow sometime in the first half of January 1731, he was a poor peasant, probably still thinking of a career in the clergy. A countryman from Kurostrov Island let him stay at his home; another lent him money.[6]

Shortly after arriving, Lomonosov was admitted to the Slavic-Greek-Latin Academy, the highest religious educational institution in Russia. A theological academy, it training members of the clergy for positions in Russian public institutions and in the Russian Orthodox Church. He was only able to pass his entrance interview by lying to its rector, Herman Koptsevich, presenting himself as the son of a Kholmogory nobleman and not subject to the poll tax levied on peasants. For some unknown reason, Lomonosov was not required to supply the usual documentation. Koptsevich might have liked Lomonosov and been anxious to recruit a bright student, or been encouraged by some Kholmogory contact to overlook the step. Koptsevich would leave the Slavic-Greek-Latin Academy later in 1731 to become bishop of Kholmogory, so he must have had contacts in that region. Whatever the reason for the shortcut, Lomonosov was duly enrolled.

A stellar student, Lomonosov was groomed for a position beyond the academy. Then came an event that nearly destroyed him, relating to expeditions that the Russian government was sending to various parts of the country. These had been conceived by Peter, one of whose desires was to assemble a better picture of Russia's geography and resources, particularly in far eastern region of Kamchatka where Russia borders the Pacific Ocean. The first Kamchatka expedition (1725–1730) was to be followed by a second (1733–1743). To accompany the latter, the Spassky Academy was asked to supply a senior student advanced enough to be consecrated before the expedition left.

Lomonosov volunteered and was chosen. He was already well-educated; had he gone, it would have meant the end of his studies and, career-wise, he would have wound up in one of the low ranks of the church hierarchy. Once again, Lomonosov was interviewed about his background to determine his suitability for the post. Forgetting his earlier fib, he asserted that his father was a priest and therefore not subject to the poll tax levied on peasants. The academy's administration had not forgotten. Lomonosov's earlier deception was exposed, he was summoned for an explanation, and he confessed he was the son of a peasant.

Not only did this make him ineligible for the Orenburg expedition, he was nearly expelled from the academy. Again it is unclear why Lomonosov escaped punishment; surely it was at least in part because he had acquired such a stellar reputation and expulsion would not have reflected well on the academy, and also because the academy was severely short on students.

Lomonosov then had three fantastically lucky breaks in a row that radically transformed both his education and his aspirations, turning him forever away from theology and sending him definitively into science. The first was set in motion in November 1735, when the Senate sent the Slavic-Greek-Latin Academy a request to send twenty of its best students to the capital to study at the St. Petersburg Academy of Sciences. Founded in 1725, the Academy had been one of Peter's Westernization projects, but it, too, was short on students. The Slavic-Greek-Latin Academy could only find twelve, Lomonosov among them. He had not formally graduated and was still in the first of the two classes in the upper level, but was clearly perfectly suited for the opportunity.

## St. Petersburg (1736)

After a decade, the Academy of Sciences was still a mill, to return to Peter's metaphor, standing on dry land. The system was not working, thus the motivation for seeking students from the Savonic-Greek-Latin Academy. This trickle of water into Peter's mill—the dozen students—left Moscow on December 23. The Academy's effective head, Johann Daniel Schumacher (1690–1761), met them when they arrived in St. Petersburg on January 1, 1736. It would be Academy university's biggest ever class.[7]

A few days later, Lomonosov moved with the other new students into the Academy's accommodations, a palace close to the Academy. He began taking the Academy's classes on experimental physics, chemistry, and mineralogy. His teachers included Vasily Adodurov (1709–1780) and Georg Wolfgang Kraft (1701–1754), a mathematician and a physicist, respectively. Lomonosov began to study a book by the poet Vasily Trediakovsky (1703–1768), also a Slavic-Greek-Latin graduate and the first Russian commoner to be educated abroad (at the Sorbonne), who became the Academy's translator in 1732. Years later, in 1745, Trediakovsky and Lomonosov would be inducted into the Academy on the same day—the first native-born Russian academicians at the Academy. Trediakovsky's book on Russian verse ignited in Lomonosov a lifelong interest in, and inspired his contributions to, the formal structures of language. Trediakovsky, Lomonosov, and a third poet and playwright who worked in St. Petersburg, Alexander Sumarokov (1717–1777, a Cadet Corps graduate), would become key figures of Russian classicism in poetry, and the three Russian poets who most strongly transformed Russian literature—the first two by developing theory and practical rules for Russian verse, and all three by their prolific writing.

But Lomonosov did not have time to settle into the Academy early in 1736 before he had a second fantastically lucky educational break: he was chosen to be sent to the West to study science. This was yet another event in his life triggered by the Russian expeditions of exploration. Late in 1735, the administrators of one of the teams on the Great Northern

Expedition had approached Korf to see if he could supply extra scientific support. The expeditions included several scientists, including the Academy astronomer Delisle, but no chemists who might know enough about mining and mineralogy to help the expedition. Korf first approached several German scientists for suggestions, among them Johann Friedrich Henckel (1678–1744), an eminent metallurgical chemist who taught mining arts and had a laboratory in the mining town of Freiberg in Germany.[8] In January 1736, Henckel wrote Korf that he did not know of anybody, but offered to train students who might then be sent on future expeditions.[9]

Russian students had studied in Germany before; in Peter the Great's time, a total of fifty-four Russian students had studied in Germany, and fifty-one more would go in the years 1726–1744. But this was a specially targeted opportunity. Korf realized the wisdom of Henckel's proposal and in February raised the issue with the Senate, asking for a supplemental appropriation. The Senate agreed and asked for three names. On March 13, Korf gave them the names of Lomonosov, Gustav Reiser, and Dmitry Vinogradov. Reiser (1718–after 1754) had been born in Moscow. His father was Vincent Reiser, an important Russian mining minister who had served for several years in Sweden as the secretary to the president of the Royal Berg College. The elder Reiser also had connections to Schumacher, and had first-hand knowledge of Saxony's mineralogists and geologists, having been sent there by the Senate to recruit twenty specialists for work in Russia.[10] Vincent Reiser's son Gustav was highly regarded as an assiduous student. Dmitry Vinogradov (1720–1758), the son of a priest, had been born in Suzdal in West-Central Russia and had attended the Slavic-Greek-Latin Academy, where he had befriended Lomonosov, who he had accompanied to St. Petersburg. Though proficiency in German was supposed to be a requirement, only Reiser knew it well. The German requirement waived for Lomonosov and Vinogradov, given that they were each proficient in Latin, were clearly adept at languages, and could take German lessons when they got there. Presumably within a short time they would be fluent enough to take classes alongside other Marburg students. Korf notified the three of his decision on March 19.

Schumacher, always on the lookout for opportunities for supplemental appropriations from the Senate, was alert enough to see this as an opportunity for additional, badly needed funding. Sending three Russian superstar students to Europe to study mining and mineralogy to prepare them for applying science to Russia's needs on great expeditions was sure to make a splash at the Imperial court. He and Korf managed to convince the Senate to allot 1,200 rubles per year of special funding for the three students. No solid study program had been drawn up for them, and it was possible that they would have to be sent to Holland nor England for additional training, so a quarter of that amount was held back in the Academy treasury for contingencies or for what we would call "overhead" today. That left 300 rubles a year annually for each student. Still, that was a huge amount; one week of three meals a day in Germany cost about 1 thaler, or 0.8 ruble. By comparison, 300 rubles was what the Academy's adjuncts received as their yearly salary (full professors received 600 rubles or more, gymnasium teachers 60 rubles, and students 10–20 rubles).

Korf soon had to rethink the plan. Asked by an official institution of the formidable Russian Empire to take on students, Henckel saw an opportunity for money and demanded an exorbitant fee to teach the three students—1,200 thalers, well over Henckel's annual salary at the time. This was insultingly greedy, and far more than the Academy was prepared to pay. Following Vincent Reiser's suggestion, Korf decided to economize by sending the three students to study first in Marburg with Christian Wolff (1679–1754), a professor of philosophy and physics at Halle who had advised Peter during the founding of the Academy. Under Wolff, the students could develop their knowledge of German, math, chemistry, and natural philosophy in preparation for their later work with Henckel. This was not as improvised as it might seem. At the time, the "mining arts" or "mining sciences" that they would be learning from Henckel required a more than passing knowledge of physics, chemistry, mathematics, metallurgy, and even law, which the students could indeed acquire in Marburg. Reiser's father helped draft the new plan of studies. Eventually, Schumacher negotiated Henckel down to about a third of his requested amount. Arrangements were duly made with Wolff, who was enthusiastic and supportive.

Korf gave the students seven points of instruction. They were to "show decent matters and behavior, and to pursue science the best they could." They were to understand that their first goal was mining and chemistry but were also to learn natural history, physics, geometry, trigonometry, mechanics, hydraulics, and hydrotechnology. They were to learn and obey Wolff (figure 3). They were to get their hands dirty in chemistry lab work and mining. They were to excel in languages, including German, Latin, and French. Every six months, they were to send back to the Academy a report of their activities. They were to understand that they might be sent for further education in Saxony, Holland, England, or France, in which case they would await further instructions. "Follow these instructions precisely!" Korf concluded.[11]

Marburg, a rather small school in a rather small town, was not the foremost German university. Its distinction was that it was relatively progressive; it had been founded in 1527 as the first specifically Protestant University in the German states (other German universities, most prominently Wittenberg where Luther taught, had become Protestant after their foundings), and its professors could deviate from Catholic dogma and were receptive to the new arts and sciences (figure 4). In 1738 it had only 122 students. There were no students from Russia before, and practically none after Lomonosov, Reiser, and Vinogradov. Out of 470 Russian students accepted in German Universities in 1698–1810, just four were sent to Marburg (the chief draw was Göttingen with 129). When a few years later, in 1751, Lomonosov would send two students to study in Germany. They went to Leipzig and Leiden.[12] Yet from Korf's perspective, Wolff's role in the establishment of the Academy, and the fact that he was only preparing the Russian students for their real scientific training in mining in Freiberg, made Marburg an attractive place to send the Russians. But the three students took advantage of the learning opportunities they found in Marburg that went far beyond mining and included foreign languages, writing styles, scholarly argument and presentation, and experimentation.

This was Lomonosov's third lucky break—that the trio's training in Germany did not work out as anticipated and that their education ended up considerably broader than the Academy planned. At the beginning of his career, Lomonosov in particular used his Marburg experience not just widely but greedily and ambitiously.

**European Education (1736–1741)**

The three students sailed from St. Petersburg to Germany on September 6, 1736.[13] But the ship was beaten back by one of the fierce storms that made travel in the Baltic Sea adventurous and often risky. Lomonosov later wrote that he "almost drowned" during the journey.[14] A week later they sailed again from St. Petersburg, but another storm forced them to land back at Kronstadt, only forty miles out of St. Petersburg. On September 23, they were finally able to leave Kronstadt for Travemünde, a port town near Lübeck in Germany, where they arrived on October 16. From Lübeck they traveled to Marburg. They arrived on November 3, 1736, and enrolled at the university three days later (figure 5). They also met Christian Wolff for the first time.

*With Wolff in Marburg*

The man to whom the Academy had entrusted their students was one of the most celebrated and provocative scholars in Germany and throughout Europe. Christian Wolff (1679–1754), a polymath, had studied mathematics, science, philosophy, and theology in Leipzig, where he had earned his doctorate in 1703. Already as a student he struck up a correspondence with Leibniz and became his Privatdozent. Leibniz was Wolff's career mentor, sponsoring his admission to the prestigious Academy of Sciences in Berlin (1711) and securing him a job as a professor of mathematics at Halle (1707), though Wolff soon switched to philosophy. Leibniz had also suggested that Peter the Great take on Wolff as an advisor in his plans for the St. Petersburg Academy, and Peter had complied. Wolff followed Leibniz in adopting a rationalist approach to all matters, emphasizing the priority of logic and deduction. Like Descartes, Wolff thought that the scientific method should resemble mathematics, with clear definitions and logical connections. Wolff inherited Leibniz's dream of a *mathesis universalis*, or universal science, which would cover not just the natural world but all human activities including art, politics, and religion, whose language was so precise that, in it, everything could be said clearly. The titles of several of Wolff's books began with the words *Vernunftige Gedanken von*, or "Rational Thoughts on," which became a catchphrase in intellectual discussions of the day for a philosophically modern rationalist approach to any field.[15]

The catchphrase, however, was also a lightning rod, attracting the ire of theologians, especially in Halle's Pietist climate. Wolff was no atheist and, like Leibniz, believed that although there were many possible worlds, God had brought into being the best of these, whose every feature had a reason for being."[16] Still, Wolff's emphasis on rationalism seemed to threaten the authority of revelation and the very relevance of faith. His opponents were galvanized by a lecture Wolff gave in 1721 on Chinese Confucianism. What they found provocative was Wolff's judgment that Confucianism, a pre-Christian religion, contained a satisfactory, sophisticated, and even rational morality despite lacking divinely revealed knowledge. Wolff's talk, in short, intimated the irrelevance of Christian revelation to morality. It was not an entirely new thought, but in the climate of that place and time proved incendiary. Theologians accusing Wolff of atheism got the ear of Frederick William I, the King of Prussia, who banished Wolff, forcing him to flee Halle and Saxony for the more tolerant atmosphere of Marburg, in Hesse.

Wolff's sudden and dramatic banishment from Halle had the predictable effect of sending his intellectual stock soaring and granting him celebrity status. As the Swedish historian Tore Frängsmyr put it, Wolff "acquired the reputation of being *the* great mathematician and philosopher of his time."[17] In truth, Wolff's attraction was less his originality—he borrowed and simplified much of his thought from Leibniz—than that he seemed to offer a safe intellectual haven to those Europeans who were attracted to the new science but were also repelled by atheism. Wolff offered a place, that is, for those who did not think they had to choose between nature as God's creation and a mechanical and mathematical way of understanding it. Moreover, in class Wolff was magnetic, casting a spell over students. As the German law professor Johann Pütter, who attended Wolff's lectures in 1738, recalled, "He did not read from a notebook, did not dictate, and did not recite, but spoke freely and with natural ease."[18] Students slavishly wrote down everything Wolff said, word for word, capturing even his asides. According to the scholar Alexandr Morozov, in the lecture notebooks of Wolff's students one can find notes such as: "Here laughed Mr. Councilor."[19]

Not everyone was so charmed. Morozov continues, "One witty contemporary of Wolff wrote, in 1740, that Wolff's desire "to reduce everything to reason's fundamentals recalls the sealed boxes or Russian dolls which are skillfully nested one inside the next, so that 'after patiently opening them one after the other, you get to the last box promising jewelry or whatever after patiently opening them one after the other, you get to the last box promising jewelry or some other present, you only find it empty.'"

Shortly after Frederick William's death in 1740, his son and successor Frederick the Great, a champion of the Enlightenment, had Wolff recalled to Halle and honored with the title of Vice Chancellor and then Chancellor, where his popularity would peak. As the French scholar Paul Hazard wrote of Wolff, "They called him The Sage, the name of philosopher not being good enough for him. Whole nations admired him. France admitted him to honorary membership of the Académie des Sciences, the highest distinction they had to bestow. The English had a number of his works translated, convincing evidence of the esteem in which they held him, coming as it did from a nation that considered itself unrivalled in the sphere of thought and philosophy."[20]

Wolff would be the teacher, supervisor, and sometimes exasperated protector of the trio of Russian students for the next three years. He proved an excellent mentor, taking a personal interest in the Russian students, going out of his way for them—for instance, traveling to Frankfurt himself to pick up their money sent by the Russian government—repeatedly rescuing them from troubles they had brought on themselves, sometimes using his own money to support them when the Russian money was late or the students had overspent their budgets.

Wolff lived in a town house on Marktgasse in the center of Marburg. The house is now a wine store, whose basement is the wine cellar from Wolff's time, where Lomonosov must have been on many an occasion. Lomonosov lodged a three-minute walk away, at Wendelgasse 2, in the dwelling of a widow named Catherine Elizabeth Zilch, where she lived with her daughter Christina Elizabeth Zilch (Lomonosov's future wife) and Catherine's younger son Johannes (figure 6).[21] By all accounts Lomonosov was a model student, initially at least, spending most of his money on books and on paying Frau Zilch for room and board. Soon, though, the students picked up the rowdy habits of their German peers and were engaging in parties and brawls. Pütter, who lived across the street on Wendelgasse, recalls being able to see into Lomonosov's room thanks to a particularly well-placed window and observed his habits, noting that he seemed to live the life of a typical student: "By the way, two young Russians were sent from the Empress to Marburg, lived near me. One of them drew my attention by the fact that I saw him enjoying his breakfast in the morning, he had few herrings and a good portion of beer. Afterwards I learned to know him more closely, and found his diligence, his judgment, and his way of thinking. This was Lomonosov, who afterwards became famous in his fatherland. Before his departure he had promised to marry a Marburg burger's daughter, and he kept his word."[22]

A fair amount, in fact, is known about the students' personal and academic lives during this period, thanks to letters of Wolff, Korf, Vinogradov, Reiser, Henckel and others.[23] In one letter from Wolff, he says that the Russian students were taking fencing, drawing, and dancing lessons, and had purchased wigs and fancy clothing. This is not as extravagant as it may sound; these were requirements of German students as part of their social acculturation (figure 7), and were also required of Russian students back home in the St. Petersburg Cadet Corps.[24]

Academically, the Russian students were shouldering a heavy load, studying arithmetic, geometry, theoretical and experimental mechanics, theoretical and experimental physics, hydrostatics, hydraulics, French, German, and drawing. Based on Lomonosov's report to the Academy of October 15, 1738, here was a typical day:

| | |
|---|---|
| 9:00–10:00 | Lessons in experimental physics |
| 10:00–11:00 | Drawing |
| 11:00–12:00 | Theoretical physics |
| Break for lunch and short rest | |
| 3:00–4:00 | Metaphysics |
| 4:00–5:00 | Logic |

Remarks by Wolff reveal that Lomonosov was acquiring proficiency in all these fields. He excelled in foreign languages, too; he would go on to write original publications in Russian, Latin, and German, and translate from French, English, Greek, and Italian. Throw in fencing and dancing lessons, as well as Lomonosov's independent work studying Russian verse and additional readings, and it is clear how busy and intense his Marburg life was. Lomonosov had brought with him Trediakovsky's *New and Brief Method* and continued to pore over it, writing comments in the margins of his copy in 1736–1737. Though generally sympathetic to Trediakovsky's reforms and disagreeing only on points of detail, Lomonosov used rude and blunt language that made it sound like they were bitter enemies.[25] This marks the beginning of ongoing theatrical but puerile fights between Lomonosov, Trediakovsky, and Sumarokov—the three giants of Russian classicism in poetry—that lasted for the rest of their lives, fights in which they were egged on by their respective patrons.

In March 1737, Wolff wrote Korf that he was satisfied with their studies and that Lomonosov and Vinogradov were speaking German. The Russian students were taking Wolff's mechanics course and were finishing classes in arithmetic and trigonometry. In June 1737 the students sent their first six-month report back to Schumacher. That fall, Lomonosov sent a letter to Korf (September 4) in German, to impress Korf, a native speaker, with his new command of the language, and included a drawing, "Cain," to impress him with his draftsmanship (figure 8).

Meanwhile, the Russian students were also studying directly with Wolff. As a personal mentor, Wolff was superb. He was an astute observer and voracious reader, and compiled an extensive library from which his students benefitted. His mind appeared to students like a cabinet with shelves full of clearly and carefully labeled stuff. Such a mind was perhaps more useful to students than a more systematic and integrated one would have been. But Lomonosov was profoundly influenced by Wolff's systematic-to-the-point-of-plodding methodical style.

He was also a good professional role model for Lomonosov, for he cared deeply about his students and went out of his way to help them—though his new Russian charges would test his limits. Like Lomonosov, Wolff came from a lower-class background and was forced to develop a technical vocabulary in the vernacular language.

But as an intellectual mentor, Wolff's impact was mixed, arising from complex debates during that time over the best way to understand nature. European science was in the midst of a revolutionary reconception of how best to do this. Through the Middle Ages, an Aristotelian-inspired framework, developed and extended by thinkers of subsequent centuries, had prevailed. In this framework, nature consisted of patterns or "forms" imposed on an undifferentiated matter, and investigating nature meant observing and describing these forms. In the seventeenth and eighteenth centuries, however, a new perspective emerged according to which the nature of matter is best investigated as if it were a matter of individual bodies or entities, whose size, shape, and motions are what gave rise to the patterns of nature. The French word for bodies was *corps* and the generic name for this new perspective was "corpuscular philosophy." But what exactly were these individual entities? Were they material or immaterial? Divisible or indivisible? Throughout Europe, several perspectives were vying for dominance, most notably Leibniz's, Descartes's, and Newton's. Leibniz viewed the basic entities of nature as immaterial and indivisible and called them monads. Descartes called them corpuscles and regarded them as infinitely divisible but material things, like little mechanical gears that filled all of space; swirls of corpuscles called vortices provided the powerful forces needed to keep the planets moving. Newton did not bother to pay attention to the gears, but focused only on the laws that governed them, making "no hypotheses" about the bodies themselves. Newton's ultimately triumphant framework had taken time to spread from England to the Continent, where advocates of the Cartesian framework warily regarded the idea of universal gravitation without a mechanical account to back it up as a restoration of occult properties in nature.

On the positive side, Wolff had developed a kind of compromise rationalist framework in which he embraced the basic thrust of corporeal philosophy.[26] Like his mentor Leibniz, Wolff was impressed by the formal, mathematical connections between things, so his work is full of definitions and deductions. Like Leibniz, he thought that there were ultimate elements of nature that were not extended or divisible—and believed that empirical knowledge can only be of properties of these monads. Most of all, Wolff impressed upon his students the need to carry out experimentation—but insisted that they carry it out methodically, laying out their findings clearly and systematically.

On the negative side, as an exponent and popularizer of Leibniz, Wolff passed on Leibniz's skepticism of Newton's revolutionary work, most notably concerning Newton's belief in action at a distance. Wolff, indeed, was a key holdout in Europe against Newtonianism; his textbook on physics, which Lomonosov translated into Russian (figure 9), has little on Newtonian mechanics.[27] The fact that Lomonosov "remained largely uninfluenced by Newton's physical theories," writes Boss, "was Wolff's worst legacy to his Russian disciple."[28]

Wolff took a paternal liking to the three Russian students, and to Lomonosov in particular. Lomonosov spent a lot of his money on books, showing maturity and good taste in his selection of the principal works on chemistry (Johann Joachim Becker, Herman Boerhaave, Georg Ernst Stahl), and acquiring most of the prolific Wolff's works.[29] But the Russian students soon began to try Wolff's patience—and Schumacher's as well—with their drinking, brawling, debts, and all-around high living.

This was not entirely the fault of the Russian students, and at least partly a function of their coming from a highly disciplined and traditional educational climate (Moscow and St. Petersburg) and being suddenly thrown into the much wilder and less regulated German university climate. But there was another, more important factor. The university had become an important part of the local economy and Marburg was now financially dependent on the students. Many students belonged to the German nobility and even as students wore "velvet camisoles, wigs, lace shirts, silk stockings and shoes with buckles."[30] Swords, carried openly, were regarded as a necessary accessory for student attire, which all but invited street fights. Students were also expected to drink to excess. A decade earlier in 1727, at the bicentenary of the University of Marburg, the annals of the university record that the festival was held without disorder or mishap and transpired safely, "except for the fact that all the glasses, bottles, tables, benches, and windows were smashed to smithereens."[31]

Wolff's arrival had drawn an additional influx of students, which, while putting a strain on university resources, offered yet another opportunity for jacking up prices. As Andrei Andreev noted, Justin-Gerard Duising, a professor of medicine and one of the Russian students' teachers, complained that, overnight, the cost of living for a student had doubled. "Students who came from Marburg from all over Germany and beyond were forced to put up with such costs.

They brought with them more money, but this only encouraged further price hikes. The number of 'foreigners'—non-Hessian students enrolled at the University of Marburg in the 1730s—ranged from 100 to 120 people per year."[32] Unscrupulous moneylenders, who charged high interest rates for easy money, added to the students' temptations. The situation grew so bad that, in 1735, the Landgrave of Hesse-Kassel forbade merchants from lending students more than five guilders (one thaler was worth about 0.8 roubles and three guilders) but these and other regulations were widely ignored and proved impossible to enforce.[33]

Vincent Reiser, who had been to the area before, warned the Academy of what the Russian students would face in Marburg and Freiberg even before the Russian students had departed for Germany. Visitors are marked as extra sources of income, he wrote Schumacher, and "everyone tries to shake down visitors."[34] The three students were also known to be heavily subsidized by the Russian Imperial Court, making them particularly inviting targets. Not only did they have huge allowances, but they also had the ability—which they had not experienced in Russia—to purchase things on credit.

Small wonder that the lavishly financed and socially inexperienced Russian students ran up huge debts fairly quickly. Though they mastered their German coursework, they did not pick up German bargaining skills. In March 1738, Wolff wrote Schumacher to ask the students to be more frugal.[35] Chastising unruly spendthrifts was right up Schumacher's alley. In May, Korf sent Wolff a letter and enclosed with it a document entitled "Instructions for Marburg Students," written by Schumacher in German, to be passed out to the trio. It asked them to stop taking fencing and dancing lessons, to stop buying fancy clothes, and to stick to their 300-ruble a year allowance.[36] But Schumacher's scolds communicated by letter were no substitute for daily supervision by mentors. On top of that, Lomonosov was proving to be stubborn and to have a fierce temper. In October 1737 he fought with another student, leading the University Senate to sentence him to three days in the university keep—but Wolff stepped in to pay the three-thaler fine needed to spring him.[37] Wolff also had the three Russians over for dinner once a week and initially they paid him directly. After their debts mounted, the arrangement was modified so that the money did not pass through the students' hands and the Academy paid Wolff directly.

But Wolff did not let the Russian students' behavior interfere with his intellectual assessments. That August, he sent Korf a letter in which, among other things, he passes judgment on the students. He singled out Lomonosov for special praise: "Mr. Lomonosov appears to have the brightest head among them, and he could learn a lot with due diligence, for he has a desire to learn and enjoys it."[38]

On October 4, 1738, Lomonosov submitted to the university his first student thesis, "Specimen Physicum de Transmutatione Corporis Solidi in Fluidum A Motu Fluidi Praeexistentis Dependente" (On the Transformation of a Solid Body into a Liquid, Depending on the Pre-Existing Fluid Motion).[39] Modeled on Wolff's methodic expositions, it consists of thirty-four short statements labeled "definitions," "clarifications," "axioms," and so forth.

In January 1739, Wolff wrote Korf that he was having to pay off debts incurred by the students, who had not received their allowance from Russia; he also told Schumacher that it was time for them to leave Marburg, because they had finished what they had set out to do and were abusing their academic freedom.[40] Korf and the members of the Chancellery were incensed at the debts that the students had run up: some 1,200 rubles (1936 thalers) on top of the lavish amount that the Academy had provided (Vinogradov's share of the debt was 899 thalers, Lomonosov's 613, and Reiser's 414). Korf sent Wolff an extra 1,400 rubles to cover the debts, severely reprimanded the trio again and ordered them to prepare to move to Freiberg. There, they would get only 150 rubles a year per person, instead of the original 300, and the Chancellery would send it to Henkel to dole out rather than giving it directly to the students.

In February 1739—the exact day is not known—the twenty-seven-year-old Lomonosov registered to marry his landlady's daughter, eighteen-years-old Elizabeth Christine Zilch, who had been born in Marburg on June 22, 1720. Normally, a wedding would follow within two or three months of the civil registration, but in this case did not, the delay perhaps due to religious obstacles: Lomonosov was Orthodox and Elizabeth Lutheran. Elizabeth almost immediately became pregnant, and that November she would give birth to a child christened Catherine Elizabeth.

In March, pleased with signs of the Russians' apparent behavioral improvement, Wolff wrote Korf that they "not only listen to good advice in every way, but live in harmony with one another," noting that "Lomonosov also begins to take a more gentle manners."[41] Wolff's special student evidently also had been hard at work, for with the letter Wolff included a copy of another work by Lomonosov, "Physical Thesis on the Difference of Mixed Bodies, Consisting in the Cohesion of Corpuscles." In it, the intellectual influence of Wolff is clear, not just the stylistic influence. Lomonosov criticizes Newton's views on gravitation as involving an "occult quality," for any interaction between bodies must require "impulse," meaning some mechanical force.[42] He, like Wolff, could not imagine that bodies could interact without direct contact. But it also exhibits some independence from Wolff, pointing to an understanding of corpuscles in which they are not imperceptible but material.

The Russian students' departure from Marburg took place that summer on July 9. Wolff asked the three to meet him at three a.m. at his house, where he had arranged for a carriage to spirit them away. This was for a good reason. Vinogradov in particular recently had been embroiled in several violent altercations with fellow students, and Wolff was afraid that, at any other time, they might run into someone interested in pursuing an unfinished fight, delaying the departure. Wolff gave each student four "louis," or "Louis d'or," minted coins then circulating in Saxony (where Freiberg was located), for travel expenses. (A louis was 6.75 grams of gold, while a ruble was about 2 grams, so 1 louis was about 5 thalers.) Lomonosov was distraught, for he was leaving behind both his mentor Wolff and his fiancée, whom he could not bring because she was pregnant and they were not yet officially married. Describing the departure to Schumacher, Wolff wrote that he was sorry to see them go, even as he complained again about their "dissolute life" and preoccupation with women and fighting. Strangely, given that he was Lomonosov's mentor, he seems not to have known about Lomonosov's engagement—in the form of the civil registration to marry—to Elizabeth. But Wolff also described the students as sorry to leave, writing that Lomonosov "could not say a word because of grief and tears."[43]

A few weeks later, Wolff wrote Korf again saying that the students apologized profusely to him for their behavior and debts, and promised to mend their ways. Though Wolff continued to express warm feelings about the trio, he and the Chancellery would spend months sorting out the debts that they had incurred.

*With Henckel in Freiberg*

With their background education behind them, the Russian students were now able to acquire their more professional training in mining arts in Freiberg. Freiberg was well known throughout Europe as a mining town. It had even once been visited by Peter the Great en route to the spa town of Carlsbad in September 1711. When he arrived at the castle, a procession of miners and smelters was staged in his honor; delighted, Peter commanded that ten barrels of wine be rolled out to them, which were promptly consumed amid enthusiastic toasts to his health.[44] On his return that October, Peter was shown the mining operations and even descended into the mine itself, taking up a hammer and crowbar to extract a few pieces of ore. After his departure, the miners saved those instruments as tokens of his visit.

The three Russian students arrived in Freiberg on July 13, 1739, and began to study practical metallurgy and mining with Henckel. Henckel held classes in a laboratory he had built (see figure 10), but also sent them to visit mines in the area—in Himmelsfürst, in Brand-Erbisdorf, about eight kilometers south of Freiberg, and to the Neue Hoffnung Gottes mine in Bräunsdorf, in the Freiberg area. A careful observer, Lomonosov not only absorbed the lessons but noted the dialect of the miners he saw, and studied their life, work, and customs as well, enough to grow concerned about their safety. Years later he would write about these experiences in his book *Foundations of Metallurgy and Ore Mining* (1763).

That August, the St. Petersburg academician Gottlob Junker (1705–1746) visited Freiberg. Junker was an itinerant scholar and poet whom Anna Ioannovna, who became Empress of Russia in 1730, had appointed professor of eloquence at the Academy in 1734; one of his first duties was to translate Trediakovsky's "Ode on the Surrender of the City of Gdansk" into German. The following year he was recruited to be a historian for Field Marshal Count Burkhard Christoph von Münnich, who was waging a campaign in the Russo-Turkish War against the Ottomans. In 1737 Münnich sent Junker to Germany to study the salt business. Two years later, on his way back to St. Petersburg, Junker stopped in Freiberg for four months, where he studied Henckel's educational program and how studiously the Russians were absorbing it (see figure 11); Lomonosov appears to have learned from Junker much about the salt business in turn. Juncker reported back to the Academy on the students. From their outward appearance they may look like "slobs," Junker wrote, but they were "diligent" and "curious" in their studies, and acquiring a "beautiful foundation" in mining science.[45]

Lomonosov also continued his studies of poetry. Starting in September 1739, after learning of Münnich's capture of the Ottoman fortress at Khotin the month before, he spent two months writing an ode called "On the Taking of Khotin," as well as an essay entitled "Letter on the Rules of Russian Poetry." These were Lomonosov's responses to Trediakovsky's "Ode on the Surrender of the City of Gdansk" and *New and Concise Way to Compose Russian Verse*, which in the standard spirit of youthful rebellion Lomonosov roundly, openly, and somewhat unfairly criticized. He sent the ode and essay back to the Academy, where they were studied by the Russian Assembly members: Adodurov, Staehlin, and Trediakovsky himself. The first two marveled at it—even Staehlin, who was more familiar with German poetry. But Trediakovsky, who viewed himself as *the* Academy's poet and linguist, was annoyed, and composed a critical reaction in turn, now lost, on the Academy's behalf. But Adodurov and Staehlin persuaded him not to send it. The reasons might have been that they were not in full agreement with Trediakovsky, that discussion of such matters by mail was not effective, or that Lomonosov was still just a student and away in Germany on a totally different mission, so why bother stirring up trouble? Such trouble could, and did, await Lomonosov's return. Whatever Trediakovsky's criticisms, Lomonosov's ode has since come to be regarded as revolutionary. It was deeper and more sophisticated than Trediakovsky's, and still sounds somewhat fluid, at least compared to previous Russian poetry, to modern Russian ears. A few later critics, such as Vissarion Belinsky in the mid-1800s, even dated modern Russian poetry from Lomonosov's "Khotin."[46]

Back in Marburg, on the evening of November 8, 1739, Elizabeth Christine Zilch gave birth to Lomonosov's daughter, Catherine Elizabeth. The archives of the St. Catherine Lutheran

Church record the child as illegitimate, due to the fact that only the intent to marry had been recorded, and that the church ceremony had not yet taken place.

In Freiberg, the Russian students encountered the same dangerous cultural environment that they had found in Marburg and promptly fell into the same temptations. Here, too, the local vendors did not miss the chance to shake down the three wealthy Russian students; as Henckel himself put it while defending his charging the Russians more than his other students, "The Russian queen is rich, and can pay even twice as much [as German students]."[47] Despite their earnest promises to Wolff the morning of their departure from Marburg, it did not take long for the Russian students, and Lomonosov in particular, to return to their previous habits of drunk and disorderly conduct.

Quarrels between Henckel and Lomonosov erupted by late 1739. In December, after one outburst, Lomonosov refused to attend classes for two days. The ostensible reason was that Henckel wanted him to titrate mercury, and Lomonosov objected to the smell and having to do menial lab work. But Henckel (see figure 12) does not seem to have been entirely innocent. Lomonosov regarded Henckel as exploiting his Russian charges both monetarily and as assistants, and accused Henckel of not apprenticing them well enough.[48] Afterwards Lomonosov wrote a letter of apology, and then met with Henckel personally to ask forgiveness.

But the antagonism between Henckel and the Russian students deepened. Henckel began sending the Chancellery reports of the students' disorderly conduct and profligate spending. The Chancellery responded by harshly reprimanding the students. In a letter to them, the Chancellery said that it was regarding Henckel's reports of the students' behavior "with great indignation," and that, although the three students would not be disciplined, they were ordered "with the utmost diligence to give themselves over to studying metallurgy," to associate only with reputable people, and to do what Henckel said.[49] Meanwhile, the Chancellery sent money directly to Henckel to pay off the students' debts.

In April 1740, Lomonosov and Heckel clashed again and Lomonosov stormed wordlessly from Henckel's lab, spitting on the way out. Lomonosov's complaints were that Henckel delayed turning money over to the Russian students; that he was exploiting their help, not mentoring them well and teaching chemistry too slowly, spreading out one month of material into four months while passing up opportunities to teach them metallurgy and mineralogy; and that he ridiculed Lomonosov's attempts to explain chemical phenomena according to the fundamental principles of mechanics and hydrodynamics that Lomonosov had learned in Marburg.

Henckel meanwhile was antagonized by the Russian students' constant demands for money. Lomonosov, he wrote the Academy, had disrespected and insulted him, showed up at the lab drunk, was obsessed by some girl in Marburg (evidently he did not know about Elizabeth either), had stolen some of Heckel's lab equipment and ripped some of Heckel's books into little pieces. Lomonosov, Heckel said, got into fights in wine-cellars, hung out with delinquents, and was abusive to townspeople who asked him to quiet down, grabbing one by the throat and threatening him. Henckel also said that Lomonosov regularly used foul language, including a piece of local filthy slang, "Hunngt Fuit" (a dialectical form of hochdeutsch "Hundsfott") that Lomonosov had picked up and was fond of shouting at those with whom he disagreed. At the end of his rope, Henckel said that even if the students wound up begging for money in the streets he would not give them more support.[50]

**Return to Marburg (1741)**

In early May 1740, after only ten months working with Henckel in Freiberg, Lomonosov decided that it was impossible for him to study there any longer. But he could not return to St. Petersburg without official authorization, whether from the Academy or from some other Russian official. Lomonosov was reluctant to seek help from Schumacher, with whom any exchange by mail would be lengthy and no doubt unpleasant. Lomonosov therefore left his belongings with Vinogradov and departed for Leipzig, seeking to find Baron Keyserlingk, the former President of the Academy and now the Russian envoy to the Elector of Saxony. Lomonosov's idea was that Keyserlingk would understand and had the authority to order Lomonosov back to St. Petersburg, allowing him to bypass Schumacher. Reaching Leipzig on May 19, Lomonosov found that Keiserlingk had left for Kassel (figure 13). Lomonosov then set out for Kassel, only to discover that Keiserlingk was not there either, having continued to Dresden. Confused, frustrated, and in serious breach of the Academy's plans for him, Lomonosov abandoned his pursuit of Keyserlingk and left for Marburg, to rejoin his fiancée and to seek to borrow money from friends so that he could try again to make the trip back to St. Petersburg.

Schumacher shortly received letters from both Lomonosov and Henckel complaining about the other's behavior. On May 10, Lomonosov wrote Schumacher bewailing Henckel's treatment of him and explaining the reasons for his departure. He defended being back in Marburg by saying that he was taking algebra, which was much needed for chemistry and physics. He begged the Academy to free him from Henckel and send him to the Harz mountains to continue studying mineralogy and mining. He promised to be more careful with money. For his part, Henckel wrote the Academy condemning Lomonosov's behavior. Although smart and quick, Henckel wrote, Lomonosov "was not of good character" and had humiliated Henckel and his family.

The Academy continued to receive complaints from Lomonosov and Henckel through the fall. In each of his letters, though, Henckel showed respect for Lomonosov's scientific acumen. In one he remarked, "In my opinion, Mr. Lomonosov is fairly well assimilated in theoretical and practical chemistry, predominantly metallurgical, and in particular in the art of assaying, as well as the art of surveying, identifying ores, ore veins, lands, rocks, salts and waters, and is able to thoroughly teach mechanics, in which he, according to the experts, is very knowledgeable."[51]

On May 26, shortly after he arrived back in Marburg, Lomonosov and Elizabeth finally had a formal marriage ceremony, this time in the Marburg church of the Reformed community (see figure 14). Nevertheless, it was one of the lowest moments in Lomonosov's life. He had fallen from being a well-supported and celebrated student to a condition of serious debt, facing the need to support a family in secret. He was even facing the possibility of jail, thanks to his violation of the instructions of the Imperial institution that was supporting him.

Borrowing money, Lomonosov embarked on another effort to seek permission to return to St. Petersburg. Again, as a state-supported student in whom the Academy had invested significant funds and for whom they had considerable hopes, he needed permission at the highest level. This time he aimed to get it from Count Alexander Golovkin, the Russian ambassador to Holland. But the quest for permission would turn into a four-month odyssey.

Lomonosov first passed through Frankfurt and Rotterdam en route to The Hague, where Golovkin was located. But Golovkin refused to order him back to St. Petersburg, saying that it was not in his authority to do so without clear instructions from the Academy. Meanwhile, as it happened, the Academy had indeed sent Keyserlingk instructions to find Lomonosov and order him back to Russia—but Keyserlingk was now unable to reach Lomonosov as both parties were on extended travel.

Lomonosov now proceeded to Amsterdam, where acquaintances from the Archangelsk network (who were themselves ignorant of the Academy's instruction) advised him not to return to St. Petersburg without the Academy's permission. Lomonosov set out to return to Marburg, but stopped along the way in Leiden, where he met Berg-Councillor Johann Andreas Cramer (1710–1777), a noted chemist and metallurgist who had recently published a textbook about assaying. Lomonosov spent the entire summer of 1740 in the Leiden area, visiting metallurgical facilities and smelting houses.

After Lomonosov left Leiden he headed to Düsseldorf, normally a two-day journey. On the way, in a story that Lomonosov recounted to Staehlin and which the latter published in his *Anekdoten*, transpired one of the more remarkable episodes in the life of this most remarkable scientist. At one point on the trip, near the fortress town of Wesel, Lomonosov encountered a recruiter from the Royal Prussian army who was treating several soldiers and recruits to a bountiful meal. The recruiting officer invited the tall, boisterous Russian stranger to join them. Never one to turn down free food and drink, Lomonosov agreed, and partook of a convivial evening. The officer kept plying Lomonosov with alcohol, all the while selling the pleasures of life in the Royal Prussian army, until Lomonosov passed out.

Lomonosov awoke the next morning still in the company of the Prussians, remembering little of the night before. He found himself wearing a red tie around his neck—an identifying mark of a member of the Royal Prussian army—and having Prussian coins in his pocket. He was informed that, the previous evening, he had agreed to enlist in the Prussian Army as a hussar—light cavalryman—for which he had accepted a deposit. Noting Lomonosov's unhappiness, the Prussian officers made sure he was carefully monitored, and they stationed him in the fortress at Wesel. He soon began planning an escape, pretending to be delighted by service. He spent several weeks in the fortress devising a plan and looking for opportunities. One evening he awoke after midnight and noticed his comrades all sleeping. He climbed out the window, crawled past sentries, swam the moat around the fortress, and climbed over the palisade, reaching the open field. In darkness, dressed in a soaking wet soldier's greatcoat, he headed for the German border of Westphalia a few miles distant, hoping to reach it before his absence was discovered. About a quarter of the way there, he heard a cannon shot indicating discovery of a deserter, and in the distance noticed a cavalryman heading his way. Lomonosov crossed the border barely in time and slept safely for a while until he rose and sought help, portraying himself—truthfully—as a poor student. Lomonosov made his way past Arnheim to Utrecht and then Amsterdam and finally to the attention of Golovkin.

Lomonosov then headed back to Marburg, stopping by Hessen and Siegen, where he visited mines and studied mining arts. He finally arrived back in Marburg in October 1740, where he moved in with Elizabeth. He then wrote a lengthy letter to Schumacher, begging to be permanently free of Henckel and to be sent to some place where he could learn mining and metallurgy in earnest, such as the Harz Mountains.

Schumacher had already ordered Lomonosov back to St. Petersburg via Keyserlingk, but now told him directly to return as early in the spring as sailing conditions permitted. In May 1741, Lomonosov left Marburg again (for the third time in two years), leaving Elizabeth pregnant again. Lomonosov headed for Lübeck and Travemünde, from where he sailed to St. Petersburg, arriving on June 8, 1741. Meanwhile, Reiser and Vinogradov remained in Freiberg, where they would stay until l744, when Henckel died. Reiser's father pulled the strings necessary to get the Senate to send the necessary funds to Freiburg to pay all the students' debts and the necessary fees to Henckel and his associates, and finally to get the students to return to Russia.

Back in St. Petersburg, Lomonosov was given a two-room apartment in the Academy's house not far from the Kunstkamer, one room for himself and another for a student, as well as a start-up fund of fifty rubles. Though Lomonosov and the other students had been promised appointments as extraordinary professors upon their return, Schumacher and the rest of the Academy's administrators were reluctant to appoint Lomonosov given the circumstances, wondering if some additional testing or training in mining might be necessary. They would not make him a professor until 1742. In the interim, they assigned him to catalogue material in the Kunstkamera's mineralogy cabinet.

En route to St. Petersburg, again according to Staehlin's memoirs, Lomonosov had a dream about his father. Lomonosov dreamed that his father had been shipwrecked on an uninhabited island in the White Sea that the two had often visited, and that his father was now dead. Lomonosov endeavored to contact his father, but learned from merchants that he had been missing since the previous fall. Unable to leave St. Petersburg, Lomonosov asked Arkhangelsk contacts to go to the island that he had dreamt about. There they discovered the body of Lomonosov's father, and buried him on the island. The closure of news about his father relieved Lomonosov, and he began to tackle his Academy work, according to Staehlin, "with new enthusiasm and extraordinary diligence."[52]

**Conclusion**

Lomonosov returned to a different place. Much had changed in the nearly five years since he had left. The Academy had recently lost almost all of its original foreign talent, including Euler (who left three days before Lomonosov's return) and Bernoulli, due at least in part to skirmishes with Schumacher over money and various intrusions. Schumacher would now become a thorn in Lomonosov's side. The Academy had also undergone several changes in leadership. Korf had been appointed Ambassador to Denmark, and in April 1740 was replaced as President by Carl Hermann von Brevern (1704–1744). Von Brevern retired after a year (April 24, 1740, to April 15, 1741), and there was no obvious replacement. Partly this was due to Imperial changes. Von Brevern's scant year as president saw several changes in Imperial leadership. Empress Anna Ioannovna died in 1740 and on her deathbed appointed her advisor Ernst Johann von Biron (1690–1772) as regent for the two-month-old Ivan IV. Three weeks later, Biron was overthrown and arrested by his rival von Münnich and replaced as regent by Ivan IV's mother, Anna Leopoldovna, who was soon deposed by the new Empress Elizabeth Petrovna, Peter the Great's daughter. Elizabeth's lover, Alexei Razumovsky, had a young brother named Kirill whom she wanted to groom for some important position and appears to have held open the post of President of the Academy for him; Kirill Razumovsky would be appointed President of the Academy on May 25, 1746, age 18, upon his return from study in Germany. During the years that the Academy's presidency was vacant, Schumacher effectively assumed the role.

Lomonosov would soon become a major force in shaping the Academy—on top of conducting research he also developed its facilities and infrastructure, including building Russia's first chemistry laboratory based on what he had seen in Freiberg and what he thought Henckel should have provided, and editing the newspaper, both the *Sankt-Peterburgskie Vedomosti* (*St. Petersburg Gazette*, the first Russian regular new outlet and major newspaper through early to mid-eighteenth century) and the *Primechaniya* (*Notes*) to it. As Pushkin would say: "Lomonosov was a great man. Between Peter I and Catherine II, he alone is an original associate of enlightenment. He created the first university. He, better to say, himself *was* our first university." Later, Lomonosov would make other discoveries, including (during the Venus transit of 1761) the conclusion that Venus had an atmosphere. In short, Lomonosov played a significant role in reshaping the Academy along the lines of what Peter had envisioned.

As many scholars have pointed out, one major reason for the rise of modern science in Europe was the rise of a distributed network of centers of research—universities, academies, institutes, and other collectives—that both collaborated and competed with each other.[53] At the beginning of the eighteenth century, Russia did not yet belong to this network. Thanks, however, to the push delivered by Peter before his death, it was moving in that direction. Peter's Academy played a key role in sustaining the momentum.

The Academy had not yet opened when Peter died in 1725—the mill, as it were, was not only on dry land, but still under construction—but it was soon completed and began to function, spurred on by other processes that Peter had set in motion. The Westernization Peter initiated continued after Anna's accession in 1730, but, as Lipski noted, "not at the hectic speed of the Petrine period." It encountered severe resistance from the nobility, who objected to the requirement that everyone must participate in state service and who initiated several coups against that requirement. To be sure, this Westernization affected only a small segment of the population, but it happened in such a way as to allow new developments in the West to "find almost immediate reflection in Russia." This Westernization process under Anna, moreover, was no longer "due chiefly to the efforts of one overpowering personality," nor driven primarily by utilitarian motives. Rather, Lipski concluded, "it was rather the result of the uncoordinated efforts of individuals, chiefly Germans."[54]

Michael D. Gordin has argued that Peter had more of a "master plan" for Russia than he is usually credited with, one that went beyond mere utilitarian and pragmatic goals. Lomonosov's career trajectory is a good example of how extensive the Petrine impact was, illustrating how different aspects of Peter's vision intersected with and reinforced each other.[55] Without the Petrine reforms, Lomonosov would surely have ended up a cleric somewhere in Russia's hinterlands. But things such as the Russian interest in mining, the extensive Western academic contacts, and the new interest in exploring Russia's regions and resources—amounting to a kind of Petrine Grand Challenge—all contributed to the events that transformed Lomonosov into Russia's first modern scientist. Lomonosov's educational trajectory is an example of all these factors at work. With his arrival back at the Academy, poised to become part of its research faculty, the water was finally beginning to flow into Peter's mill.

## Acknowledgements

Many people were consulted on the Lomonosov's years in Moscow and Germany. We were extraordinarily fortunate to have Michael D. Gordin give us not only help with many details but also substantive suggestions regarding our overall approach. We also greatly appreciate help and advices by Dr. Peter Hoffmann and Dr. Norbert Nail—Lomonosov scholars in Germany and experts on his connections with Wolff—as well as detail explanations given Prof. Dr. Ulrike Wagner-Rau and Dr. Tobias Braune-Krickau of the Marburh University Church on the nature and tradition of interfaith marriages in the mid-eighteenth century Marburg. One of us (VS) had a memorable visit to Freiberg in summer 2017 and wishes to express sincere gratitude to the people whom he met there—Prof. Dr. Friedrich Naumann, Mr. Evgenii Kaschenko, Mrs. Tatziana Piliptsevich and Dr. Dmitry Parkhomchuk—who carefully maintain Lomonosov's legacy in Saxony and helped to make the trip so successful and very useful for this publication. Staff of the Library of *Moskovskoe Obschestvo Ispitatelei Prirodi* (Moscow Society of Nature Philosophers) Irina Prokhorova (Director) and Irina Zakharova were extremely cooperative in helping us to find the 1741 hand-written notebooks of Dmitry Vinogradov taken while in Henckel's lab in Freiberg. Finally, we also appreciate the care and patience with which Joseph D. Martin shepherded the manuscript through the editorial process.

## References

1 Vasily N. Tatishchev, "Razgovor Dvuh Priyatelei o Pol'ze Nauk i Uchilisch (The Conversation of Two Friends about the Benefits of Sciences and Academies)," in *Collected Works*, 8 vols. (1962–1971; Moscow: Ladomir, 1996), 8:51.

2 Michael D. Gordin, "The Importation of Being Earnest: The Early St. Petersburg Academy of Sciences," *Isis* **91**, no. 1 (2000), 1–31.

3 For more on Lomonosov, see Boris N. Menshutkin, *Russia's Lomonosov* (Princeton: Princeton University Press, 1952); Peter Hoffman, "Michail Vasil'evič Lomonosov (1711–1765)," in *Ein Enzyklopädist im Zeitalter der Aufklärung* (Essen: Peter Lang, 2011).

4 The principal source on Lomonosov's life is Valentin L. Chenakal, ed., *Letopis' zhizni i tvorchestva M. V. Lomonosova* [*Chronicles of the Life and Works of M. V. Lomonosov*] (Moscow: Soviet Academy of Sciences, 1961).

5 Robert P. Crease, and Vladimir Shiltsev, "Pomor Polymath: The Upbringing of Mikhail Vasilyevich Lomonosov, 1711–1730," *Physics in Perspective* **15** no. 4 (2013): 391–414.

6 Chenakal, ed., *Chronicles* (ref. 4), 22.

7 The Academy's university ceased in 1766, a year after Lomonosov's death, though it was resurrected in 1819 as the St. Petersburg Imperial University, which still exists. The Academy gymnasium was housed from 1726 until 1764 in the building next to the Kunstkamer; in 1764 it moved to a separate building. The gymnasium closed in 1805, with its students moving to the new St. Petersburg gubernia gymnasium.

[8] Henkel had an international reputation at the time, as the author of *Pyritologia oder Kies-Historie* (1725) and as a member of the Berlin Academy and the Leopoldina Academy.

[9] Johann Friedrich Henckel, letter to Johann Albrecht Korf, January 21, 1736, in A. Kunik, ed., *Sbornik Materialov po Istorii Akademii Nauk* [*Materials on the History of the Academy of Sciences*] (St. Petersburg, 1865), 225–26.

[10] J. Staehlin, *Podlinnie Anekdoti o Petre Velikom, Sobrannie Yakovom Shtelinim* (*Genuine Anecdotes from the Life of Peter the Great)* (Moscow: Reshetnikov Print House, 1830), 166–67.

[11] A. Kunik, ed., *Materials on the History* (ref. 9), 246–47.

[12] A. Andreev, *Russkie Studenti v Nemetskih Universitetah XVIII – Pervoi Polivini XIX veka* [*Russian Students in German Universities of the Eighteenth and the First Half of the Nineteenth Century*] (Moscow: Znak, 2005).

[13] On Lomonosov in Marburg, see Barbara Karhoff, "Die Studienjahre M. V. Lomonosovs in Marburg (1736–1739)," in *M. W. Lomonossow—streitbarer Wissenschaftler und Patriot*, ed. Andreas Förster and Ch. Titel, 45–54 (Berlin: Damu, 2012).

[14] Chenakal, ed., *Chronicles* (ref. 4), 34; also Mikhail Vasilievich Lomonosov, *Polnoe Sobranie Sochinenii* [*Complete Works*], 11 vols., ed. S. Vavilov and T. Kravets (Moscow: Soviet Academy of Sciences, 1950–1983), 10:479. Lomonosov's complete works are available online from Russian National Fundamental Electronic Library website: http://feb-web.ru/feb/lomonos/default.asp.

[15] Tore Frängsmyr, "Christian Wolff's Mathematical Method and its Impact on the Eighteenth Century," *Journal of the History of Ideas* **36**, no. 4 (1975), 653–68.

[16] Friedrich Engels once skewered Wolff as the proponent of a "shallow theology" in which whatever exists does so for man's good, which justified the appropriateness of natural arrangements in such a way that "cats were created to eat mice, mice to be eaten by cats and nature as a whole to testify to the wisdom of the Creator." Friedrich Engels, "Dialectic of Nature," in Karl Marx and Friedrich Engels, *On Religion*, 152–93 (Mineola: Dover, 2008), 158.

[17] Tore Frängsmyr, "Christian Wolff's Mathematical Method and its Impact on the Eighteenth Century," *Journal of the History of Ideas* **36**, no. 4 (1975), 653–68.

[18] A. Morozov, *Lomonosov* (Moscow: Molodaya Gvardiya, 1961), 164.

[19] Morozov, *Lomonosov* (ref. 18), 161.

[20] From Paul Hazard, *La pensee européene au XVIII siècle: De Montesquieu a Lessing, Tome I* (Paris: Boivin, 1946), 40.

[21] On Zilch, see Tatiana Butorina, "Lichnaya Sud'ba M.V.Lomonosova [Personal Circumstances of M. V. Lomonosov]," *Nauka I Zhizn* **11** (2011).

[22] Johann Pütter, *Selbstbiographie: zur dankbaren Jubelfeier*, vol. 1 (Göttingen 1792), 27.

[23] Wolff's own letters are collected in *Briefe von Christian Wolff aus den Jahren 1719–1753* (St. Petersburg: Eggers, 1860); letters from Wolff and the others are collected in Kunik, *Materials on the History* (ref. 9). Other important sources are Wilhelm A. Eckhardt "Lomonosow in Marburg," in *Miszellen und Vorträge,* Beiträge zur hessischen Geschichte 10 (Marburg: Trautvetter und Fischer Nachf, 1995); Hoffmann, "Lomonosov" (ref. 3), 298.

[24] Alexander Lipski, "The Foundation of the Russian Academy of Sciences," *Isis* **44**, no. 4 (1953) 349–54, on 349.

[25] Irina Reyfman, *Vasilii Trediakovsky: The Fool of the "New" Russian Literature* (Stanford: Stanford University Press, 1990), 58.

[26] For more on Wolff's views, see Thomas Broman, "*Metaphysics for an Enlightened Public: The Controversy over Monads in Germany, 1746–1748*," *Isis* **103**, no. 1 (2012): 1–23.

[27] Ronald S. Calinger, "The Newtonian-Wolffian Controversy: 1740–1759," *Journal of the History of Ideas* **30**, no. 3 (1969), 319–30.

[28] Valentin Boss, *Newton and Russia: The Early Influence, 1698–1796* (Cambridge, MA: Harvard University Press, 1972), 164.

[29] On Lomonosov's books see I. M. Belyaeva, ed., *Biblioteka M. V. Lomonosova. Nauchnoe Opisanie Rukopisei I Pechatnih Knig* [*Library of M. V. Lomonosov. Scientific Review of Manuscripts and Printed Books*] (Moscow: Lomonosov Publisher, 2010).

[30] Kunik, *Materials on the History* (ref. 9), 278.

[31] Morozov, *Lomonosov* (ref. 18), 157.

[32] Andreev, *Russian Students* (ref. 12).

[33] Andreev, *Russian Students* (ref. 12).

[34] Kunik, *Materials on the History* (ref. 9), 244.

[35] Christian Wolff, letter to Johann Daniel Schumacher, March 30, 1738, in Wolff, *Briefe* (ref. 43), 107.

[36] Kunik, *Materials on the History* (ref. 9), 270–71

[37] V. Chenakal, "Russkie Studenti iz Sankt-Peterburga: Novie Materiali iz Nemetskih Arhivov [Russian Student from St. Petersburg: New Materials from German Archives]," *Ogonyok*, November 19, 1961.

[38] Christian Wolff, letter to Johann Daniel Schumacher, August 17, 1738, in Wolff, *Briefe* (ref. 43), 109.

[39] Mikhail Vasilievich Lomonosov, *Polnoe Sobranie Sochinenii* [*Complete Works*], 11 vols., ed. S. Vavilov and T. Kravets (Moscow: Soviet Academy of Sciences, 1950–1983), 1:21.

[40] Archive of the USSR Academy of Sciences, f. 1, op. 3, No. 23, l. 226; also in Kunik, *Sbornik Materialov* (ref. 29), 281.

[41] Kunik, *Materials on the History* (ref. 9), 287.

[42] Lomonosov, *Complete Works* (ref. 39), 1:23–63.

[43] Christian Wolff, letter to Johann Albrecht Korf, August 1, 1739, in Kunik, *Materials on the History* (ref. 9), 305.

[44] Morozov, *Lomonosov* (ref. 18), 176–77.

[45] G. Juncker, letter to Korf, July 31, 1739, in Kunik, *Materials on the History* (ref. 9), 314–15. Also in F. Naumann, *M. W. Lomonossow in Freiberg* (Technische Universitat Berrgakademie Freiberg, 2011).

[46] Vissarion G. Belinskiy, *Polnoe Sobranie Sochinenii* [*Complete Works*] (Moscow: USSR Academy of Sciences, 1953), 1:65.

[47] Mikhail V. Lomonosov, letter to Johann Daniel Schumacher (from Marburg), November 16, 1740, in Kunik, *Materials on the History* (ref. 9), 335.

[48] Mikhail V. Lomonosov, letter to Johann Daniel Schumacher (from Marburg), November 16, 1740, in Kunik, *Materials on the History* (ref. 9), 334–38.

[49] Instruction of Academic Chancellery to Freiberg students, March 14, 1740, in Kunik, *Materials on the History* (ref. 9), 319–21.

[50] Mikhail V. Lomonosov, letter to Johann Daniel Schumacher (from Marburg), November 16, 1740, in Kunik, *Materials on the History* (ref. 9), 336; also in Lomonosov, *Complete Works* (ref. 39), 10:424, 429.

[51] Johann Friedrich Henckel, letter to the St. Petersburg Academy of Sciences, September 23, 1740, in Kunik, *Materials on the History* (ref. 9), 330–31.

[52] Jacob von Staehlin, *Cherty i Anekdoty dlya Biografii Lomonosova, Vzyatie s Ego Sobstevennih Slov Shtelinim. 1783* [*Features and Anecdotes for the Biography of Lomonosov, Taken from His Own Words by Shtelin, 1783*]; Publication and commentary in G. E. Pavlova, *M. V. Lomonosov v Vospominaniyah i Harakteristikah Sovremennikov* [*M. V. Lomonosov in Memoirs and Characteristics by Contemporaries*] (Moscow-Leningrad: Academy of Sciences of the USSR, 1962), 51–60.

[53] Robert Wuthnow, "The Institutionalization of Science," in *Meaning and Moral Order*, 263–98 (Berkeley: University of California Press, 1987).

[54] Lipski, "Foundation of the Russian Academy" (ref. 24), 10–11.

[55] Gordin, "Importation of Being Earnest" (ref. 2).

# Figures:

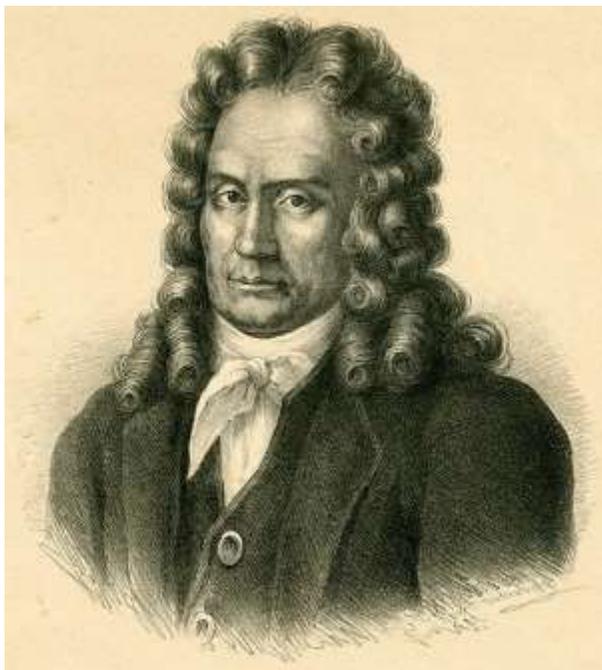

Fig.1: Lawrence (Luarentius, Lavrentii) Blumentrost (1692-1755), first President of the St.Petersburg Academy of Sciences.

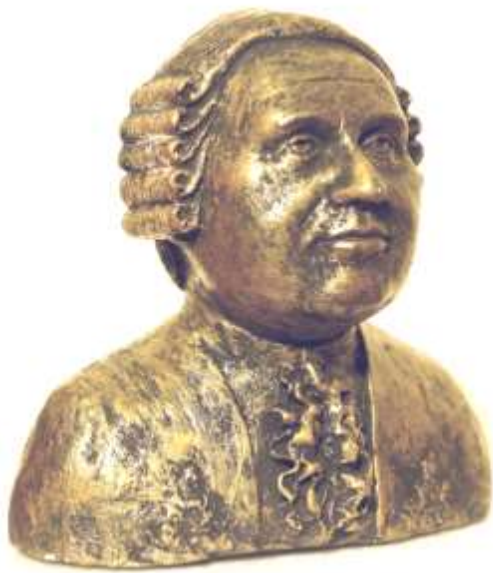

Fig.2: Mikhail Vasilievich Lomonosov (1711-1765). Sculpure by A.L.Zunde, artificial stone (2015), private collection.

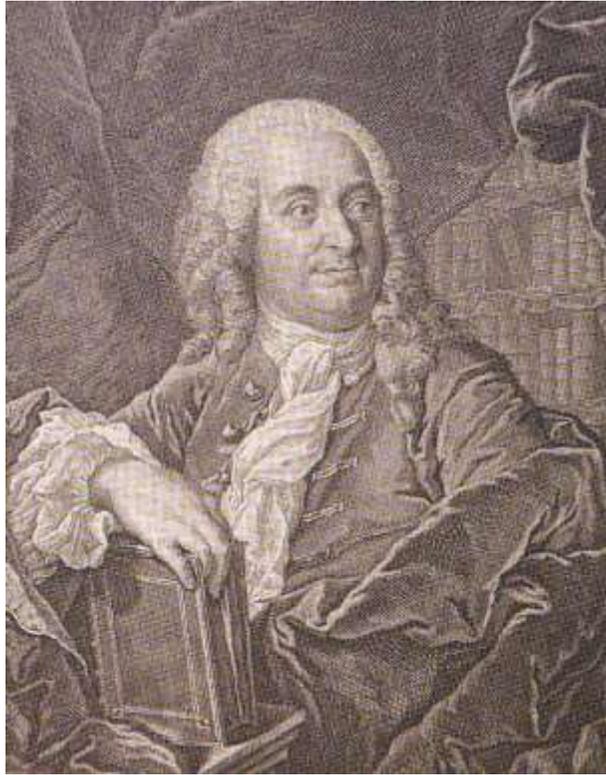

Fig.3: Christian Wolff (1679-1754), Lomonosov's professor in Marburg University (enrgaving by J.G.Wille, mid-XVIII century).

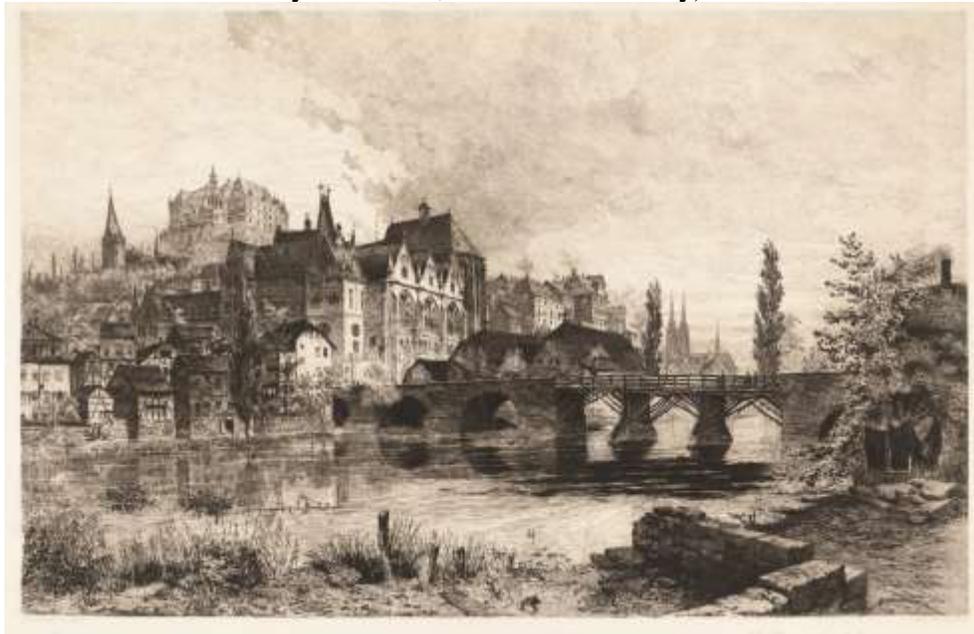

Fig.4: Philipps University in Marburg, view from the Lahn, engraving by by Ludwig Dihm (1849-1928)

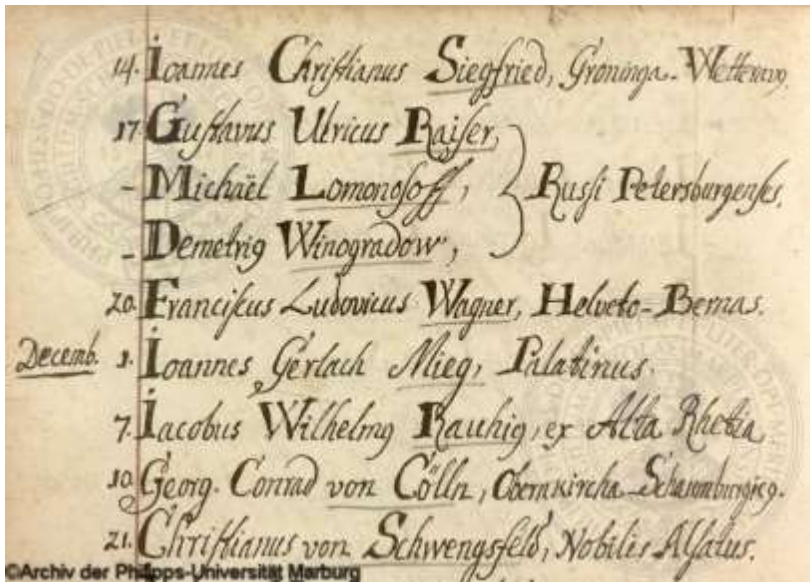

Fig.5: Excerpt from the Marburg University enrollement book, listing three "Russi Petersburgenses" - Gustavus Ulricus Reiser, Michael Lomonosoff, Dementviy Winogradow (lines 17-19, 1737).

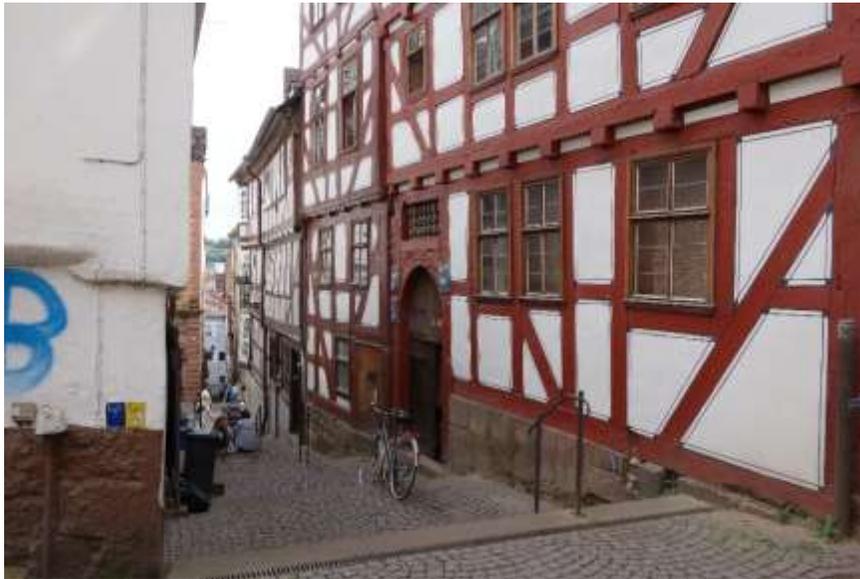

Fig.6: Modern-day photo of Lomonosov's residence in Marburg at Wendelgasse, 2.

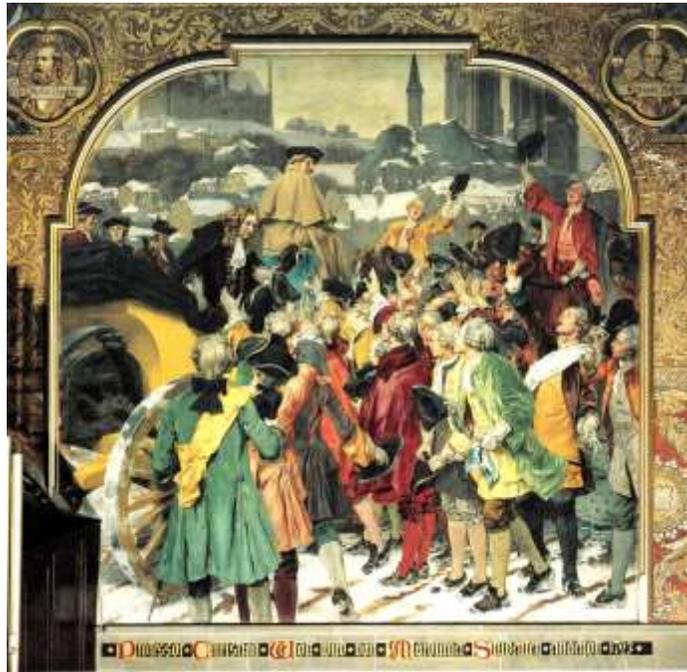

Fig.7: 1903 painting by Peter Janssen (1844-1908) in the auditorium of the Marburg Philipps University "Professor Christian Wolff welcomed by Marburg students. 1723." (reprinted from M.Lemberg, G.Oberlik, *Die Wandgemälde von Peter Janssen in der Alten Aula der Philipps-Universität zu Marburg*. (Marburg: N. G. Elwert Verlag, 1985), by permission of Gerhard Oberlik, Marburg/Germany).

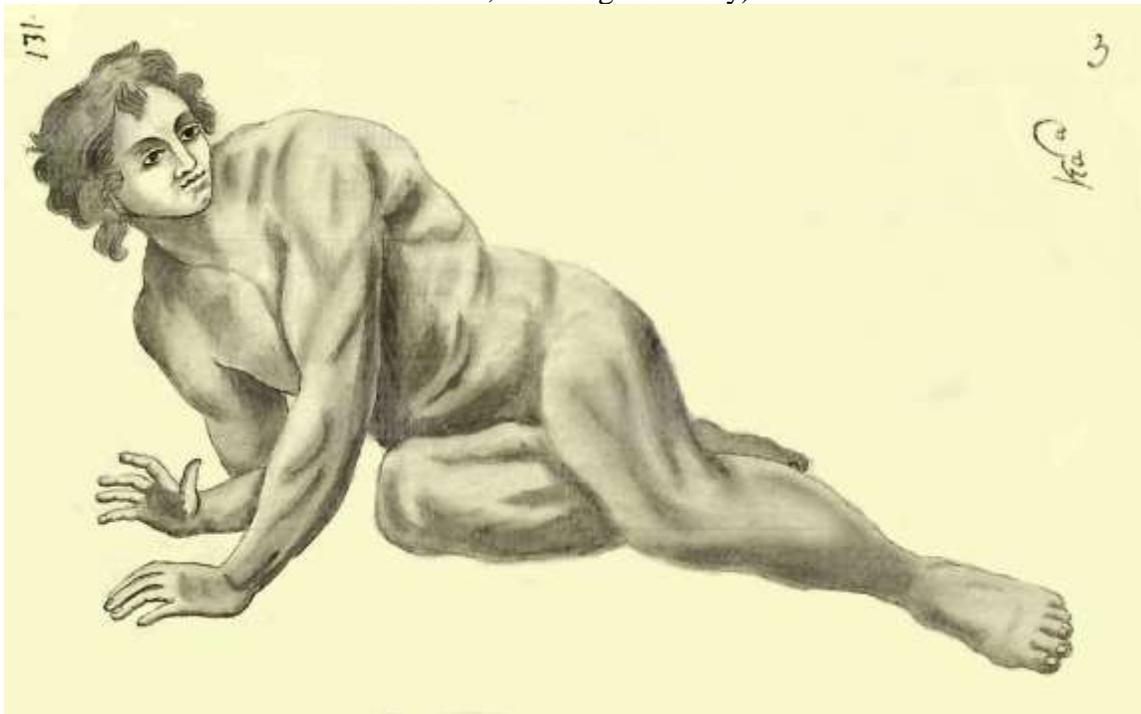

Fig.8: "Cain", Lomonosov's drawing sent from Marburg to the Academy as a proof of his progress in education.

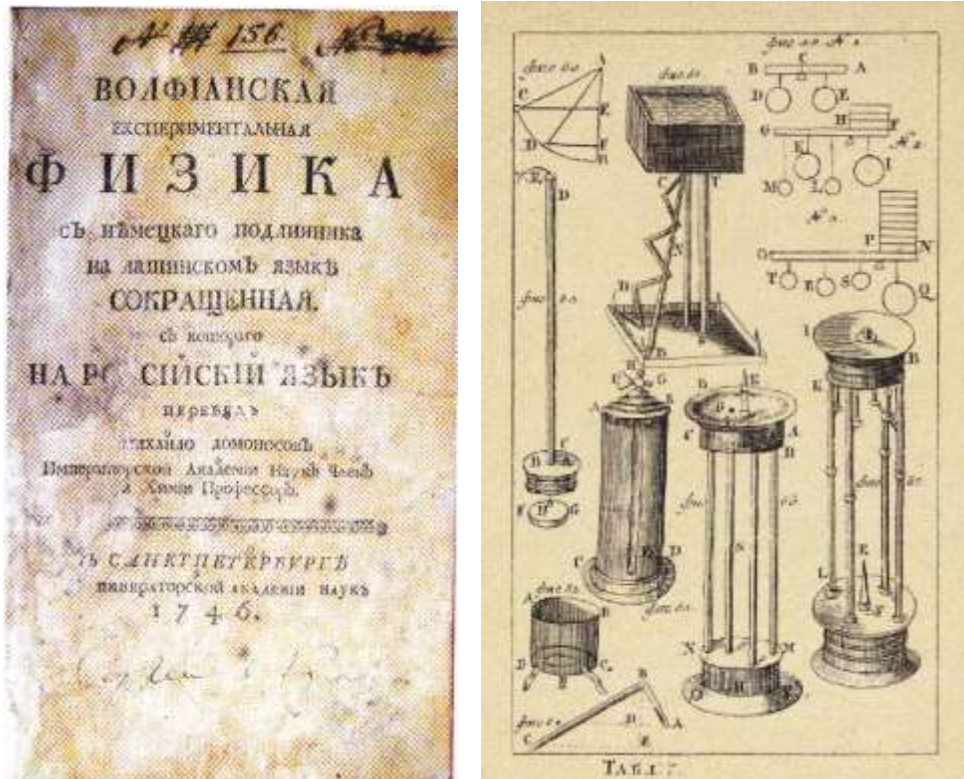

Fig.9: "Wolffian Experimental Physics" – textbook translation to Russian by Lomonosov (1746), printed by St.Petersburg Imperial Academy of Sciences (front page and one of the engraving plates).

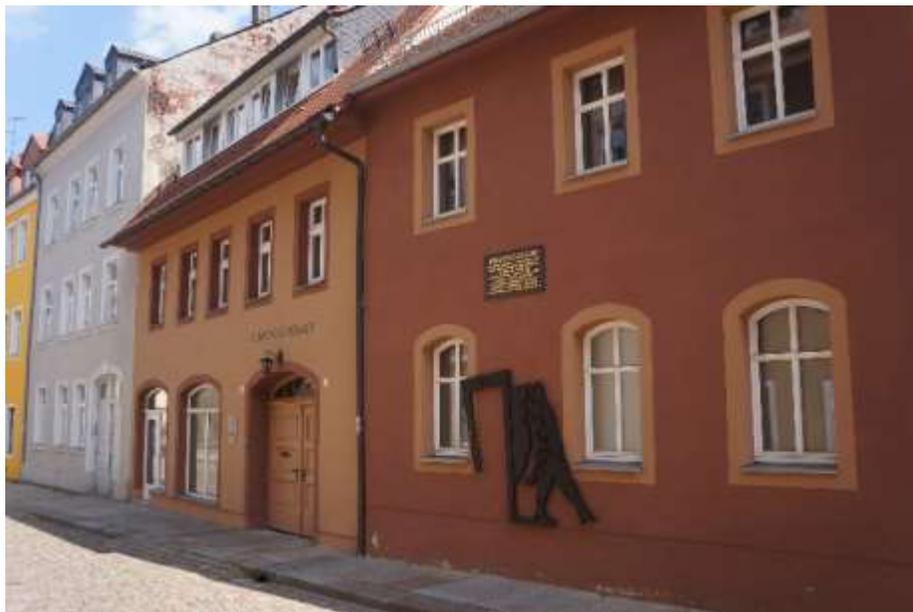

Fig.10: "Lomonossow Haus" in Freiberg - memorial to Mikhail Lomonosov, the place of Henckel's lab attended by Lomonosov in 1739 at Fischerstrasse, 39-41.

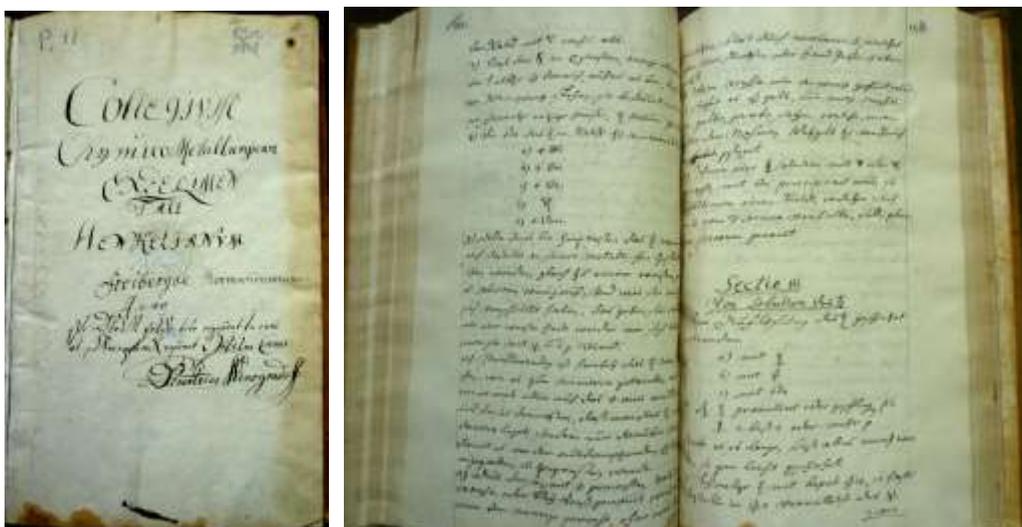

Fig.11: Handwritten notes of Henckel's lectures taken by Dmitry Vinogradov in 1741. 225 pages volume entitled "Collegium chymico-metallurgicum experimentale Henkelianum freibergae. Hermundurorum. Anno… Demetrius Winogradoff" was recently discovered in the library of the Moscow Society of Nature Philosophers (*Moskovskoe Obschestvo Ispitatelei Prirodi*).

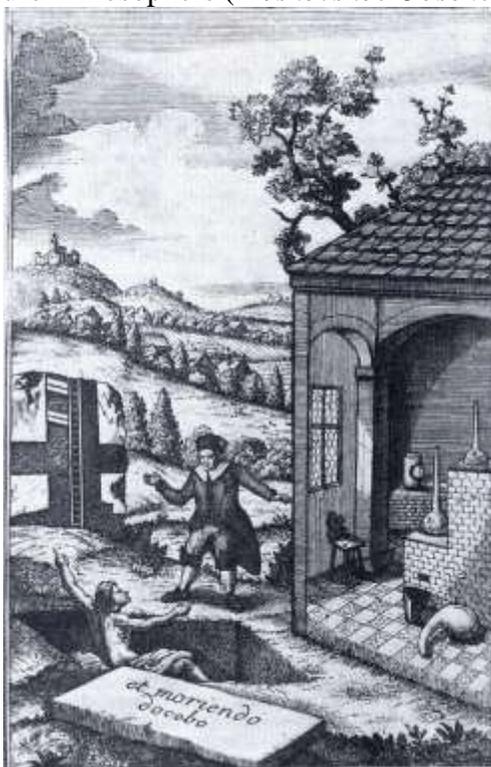

Fig.12: No portraits of Johann Henckel (1669–1744) survived. This engraving depicts him proclaiming from his grave "*Et Moriendo Docebo*!" ("*Dying but teaching*!"). Henckel's chemical lab is shown on the right (from J.F.Henckel, Johann Emanuel Stephani, *Henkelius in mineralogia redivivus, das ist: Hencklischer aufrichtig und gründlicher Unterricht von der Mineralogie oder Wissenschafft von Wassern, Erdsäfften, Saltzen, Erden, Steinen und Ertzten : nebst angefügten Unterricht von der chymia metallurgica.* Dresden: Joh.Nicol.Gerlach, 1747)

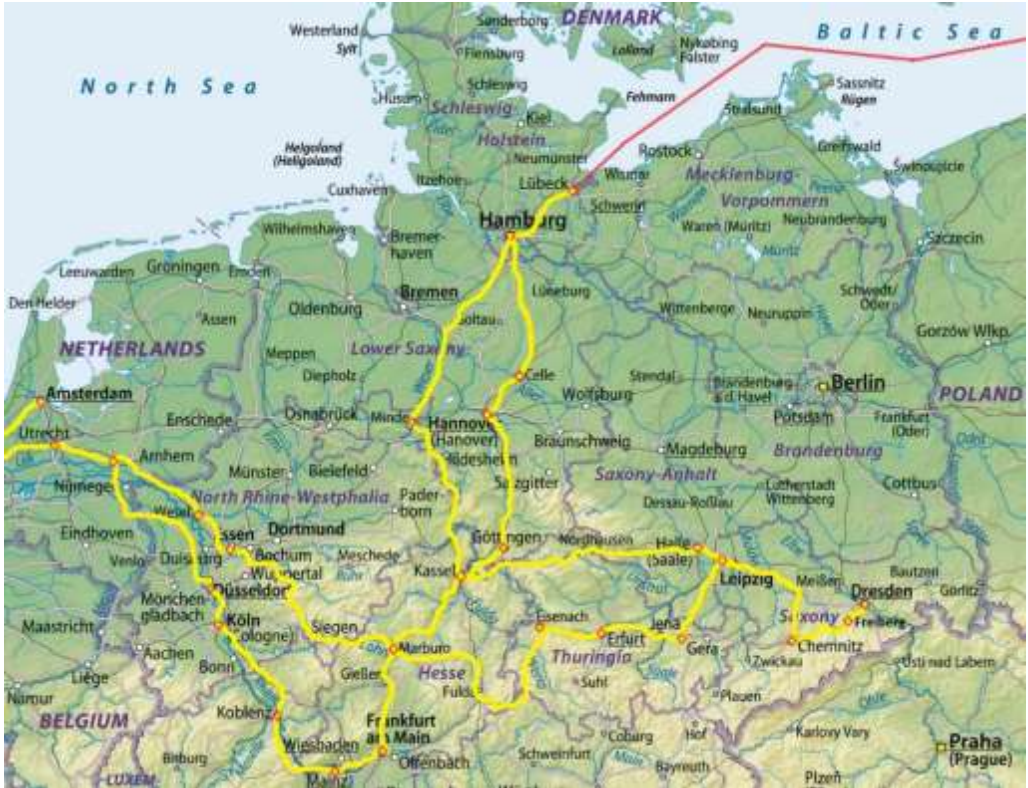

Fig.13: Lomonosov routes in Germany and Netherlands in 1736-1741.

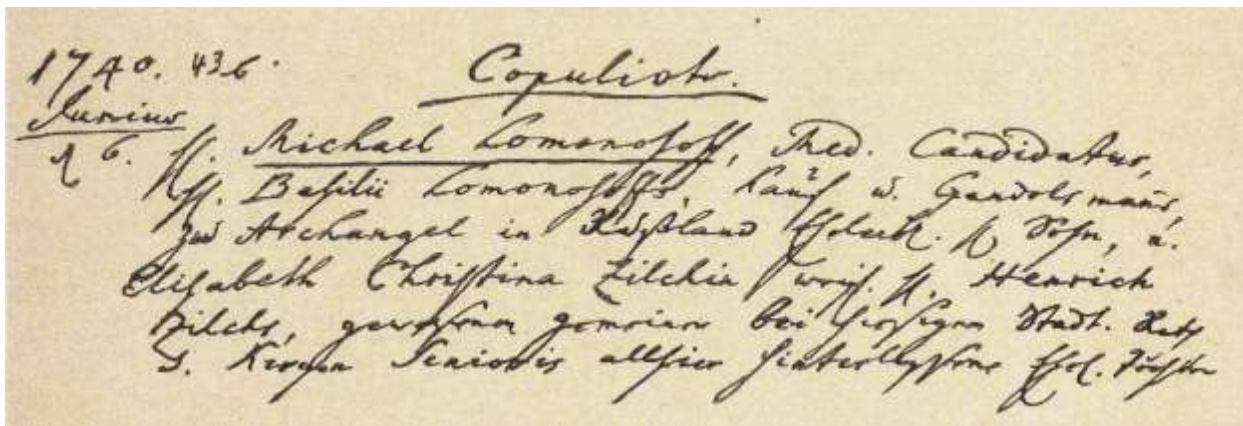

Fig.14: Entry in the record book of the Reformed Church on the marriage of Mikhail Lomonosov and Elisabeth Zilch on 6 June 1740: "*Herr Michael Lomonosoff, med.candidatus, H. Basilii Lomonosoffs, Kauf- und Handelsmannes zu Archangel in Russland eheleiblicher H. Sohn, und Elisabeth Christine Zilchin, weylands H. Henrich Zilchs, gewesenen Gemeinen bei hiesignem Stadtrath und Kirchen-Seniors allhier hinterlassene eheliche Tochter*". (Married are Mr. Michael Lomonosoff, medical candidate, son of Basilii Lomonosoff, merchant and tradesman in Archangel in Russia, and Elizabeth Christine Zilch, daughter of late Henrich Zilch, who was a town councilor and a church senator.)